\documentclass[referee]{aa} 

\usepackage{amsmath, amssymb}

\usepackage{times}

\usepackage[dvips]{graphicx}
\usepackage{epsfig}
\usepackage{txfonts}
\usepackage{natbib}

\begin{document}

\title{High-energy pulses and phase-resolved spectra by inverse
  Compton emission in the pulsar striped wind - Application to Geminga.}

\author{J\'er\^ome P\'etri \inst{1,2}}

\offprints{J. P\'etri}

\institute{Observatoire Astronomique de Strasbourg, 11 rue de
  l'Universit\'e, 67000 Strasbourg, France. \and Max-Planck-Institut
  f\"ur Kernphysik, Saupfercheckweg 1, 69117 Heidelberg, Germany.}

\date{Received / Accepted}

\titlerunning{Pulsed IC emission in the pulsar striped wind}

\authorrunning{P\'etri}

\abstract
{Although discovered 40~years ago, the emission mechanism responsible
  for the observed pulsar radiation remains unclear. However, the
  high-energy pulsed emission is usually explained in the framework of
  either the polar cap or the outer gap model.  Here we explore an
  alternative model based on the striped wind.}
{The purpose of this work is to study the pulsed component, that is
  the light-curves as well as the spectra of the high-energy emission,
  above 10~MeV, emanating from the striped wind model. Gamma rays are
  produced by scattering off the soft cosmic microwave background
  photons on the ultrarelativistic leptons flowing in the current
  sheets. }
{We compute the time-dependent inverse Compton emissivity of the wind,
  in the Thomson regime, by performing three-dimensional numerical
  integration in space over the whole striped wind. The
  phase-dependent spectral variability is then calculated as well as
  the change in pulse shape when going from the lowest to the highest
  energies.}
{Several light curves and spectra of inverse Compton radiation with
  phase resolved dependence are presented. We apply our model to the
  well-known gamma-ray pulsar Geminga. We are able to fit the EGRET
  spectra between 10~MeV and 10~GeV as well as the light curve above
  100~MeV with good accuracy.}
{In the striped wind model, the pulses are a direct consequence of the
  relativistic beaming effect. It is a simple geometrical model
  capable to explain the very high-energy variability of the
  phase-resolved spectrum as well as the shape of the associated
  pulses. Future confrontations with observations at the highest energies and
  possibly polarization measurement will be important to discriminate
  between existing models.}
   
\keywords{Pulsars: individual: Geminga -
  Radiation mechanisms: non-thermal - Gamma rays: observations - Gamma
  rays: theory - Stars: winds, outflows }

\maketitle

\section{INTRODUCTION}

To date, seven pulsars are known to radiate most of their energy in
the gamma-ray band (about a few GeV), with high-confidence level
\citep{2003astro.ph.12272T}. This high energy radiation is probably
related to some ultrarelativistic particles which are flowing in the
vicinity of the neutron star.  Nevertheless, the detailed origin of
the pulsed emission of a pulsar remains still poorly understood. A
self-consistent description of the particle acceleration in the pulsar
magnetosphere has also not been proposed yet. Many attempts have been
suggested to explain the radiation process from a strongly magnetized
rotating neutron star. Among these, the polar caps
\citep{1975ApJ...196...51R} and outer gaps \citep{1986ApJ...300..500C}
are the most extensively studied models. The extremely stable time of
arrival of the pulses supports the idea that the emission is probably
coming from regions close to the neutron star surface.  However, how
close it should be is not clear yet.

In the present work, we focus on an alternative model, the so-called
striped wind model originally introduced by
\citet{1990ApJ...349..538C} and \citet{1994ApJ...431..397M}. In this
picture, emission is expected to occur outside the light cylinder but
still in regions where the striped structure is well defined and
rotates at the angular speed of the compact object.  We use an
explicit asymptotic solution for the large-scale field structure
related to the oblique split monopole and valid for the case of an
ultrarelativistic plasma \citep{1999A&A...349.1017B}. Due to
relativistic beaming effects, it has already been shown that this
model gives rise to pulsed emission~\citep{2002A&A...388L..29K} and
can satisfactorily fit the optical polarization data from for instance
the Crab pulsar~\citep{2005ApJ...627L..37P}.

It is usually claimed that the precise shape of the cut off of the
pulsed spectrum at the highest energies ($E_\gamma>10$~MeV) will help
to discriminate between different existing models.  The outer gap
scenario predicts an exponential cut-off, \cite{1996ApJ...470..469R}
whereas the polar cap model predicts a super exponential cut-off,
\cite{1996ApJ...458..278D}. In the near future, the phase-resolved
spectra measured by GLAST will be a valuable tool to confront emission
models to observations at the highest energies (20~MeV-300~GeV)
\citep{2007arXiv0711.4278T}. The recent discovery of pulsed emission
above 25~GeV in the Crab will put some severe constraint on emission
models \citep{2008arXiv0809.2998T}.

Emission mechanisms outside the light-cylinder have a long history, as
old as the discovery of the first pulsar. Indeed,
\cite{1969Natur.223..277M} proposed a supersonic magnetically
dominated wind launched from the rotating neutron star, that
eventually enters a shock and becomes subsonic.
\cite{1970ApJ...159..229L} assumed that the low-frequency, large
amplitude electromagnetic wave propagates in vacuum and is separated
from the interstellar medium by an infinitely thin current sheet.
This current sheet oscillates due to some magnetic pressure modulation
imposed by the rotation of the pulsar. He also estimated the produced
pulsed emission. In \cite{1970ApJ...160.1003L, 1970ApJ...162..153L},
he extended this idea to include finite Larmor radius effects which
means a finite thickness of these current sheets.
\cite{1976SvA....20..299D} studied a similar phenomenon, namely the
interaction between the spiral magnetic field structure and the
interstellar medium in order to produce coherent emission.  Coherent
pulsed radiation emanating from a relativistically expanding sheet of
charges flowing along dipolar magnetic field lines has been deeply
investigated by \cite{1971Ap&SS..12..193T, 1971ApL.....8..167G}.
\cite{1975Ap&SS..33..111F} found that pulsed radiation is possible at
the light-cylinder and computed the associated synchro-compton
spectra. The formation of neutral sheets and radiation in the wave
zone is exposed in \cite{1971CoASP...3...80M}. The pulsed high-energy
emission could also be explained by magnetic reconnection (leading to
relativistic heating of particles) in the current sheets just beyond
the light-cylinder, see \cite{1996A&A...311..172L}.
\cite{2000MNRAS.313..504B} used inverse Compton scattering of the
unshocked wind on low-frequency photons coming from the pulsar to
compute the pulsed and unpulsed component of the gamma ray emission.
However, for some criticism about pulsed radiation coming from outside
the light-cylinder and related beaming effects, see
\cite{1979SSRv...24..437A}.

Attempts to explain the pulse profiles and phase-resolved spectra of
the Geminga pulsar have already been investigated by
\cite{2001MNRAS.320..477Z} in the framework of a thick outer gap
model. They used the data analysis performed by
\cite{1998ApJ...494..734F} according to EGRET observations. However,
in their study, due to the geometry of the radiating region, no
emission is expected in some phase of the rotation period of the
neutron star. For instance, there is no leading wing before the first
pulse or no trailing wing after the second pulse. They do not propose
a fit of the light-curve at high energy. We will show that our model
can explain both, the light-curve and the phase-resolved spectra.

The paper is organized as follows.  In Sec.~\ref{sec:Model}, we
describe the model, i.e. the magnetic field structure adopted, based
on the asymptotic solution of \cite{1999A&A...349.1017B}. The
properties of the emitting particles in the striped wind, and the
inverse Compton emission process are also discussed in this section.
Next, in Sec.~\ref{sec:Application}, we present the phase-resolved
spectral variability and the change in the pulse profile of the
light-curves with respect to photon energy. These results are then
applied to fit the EGRET data ($E_\gamma>10$~MeV) for one gamma-ray
pulsar, namely Geminga which is presumably not inside a synchrotron
nebula.  The conclusions and the possible extension are presented in
Sec.~\ref{sec:Conclusion}.

\section{THE STRIPED WIND MODEL}
\label{sec:Model}

The model used to compute the high-energy pulse shape and the
phase-resolved spectrum arising from the striped wind is briefly
presented in this section. The geometrical configuration is as
follows.  The magnetized neutron star is rotating at an angular
speed~$\Omega_*$ directed along the $(Oz)$-axis i.e. the rotation axis
is $\vec{\Omega}_*= \Omega_* \, \vec{e}_z$. We use a Cartesian
coordinate system with coordinates~$(x,y,z)$ and orthonormal
basis~$(\vec{e}_x, \vec{e}_y, \vec{e}_z)$. The stellar magnetic
moment~$\vec{m}$, assumed to be dipolar, makes an angle~$\chi$ with
respect to the rotation axis, $\chi=\hat{(\vec{\Omega}_*,\vec{m})}$.
This angle is therefore defined by $\cos\chi = \vec{m} \cdot \vec{e}_z
/ m$ where $m=||\vec{m}||$ is the magnitude of the stellar magnetic
moment. Moreover, it rotates around $\vec{e}_z$ at a speed $\Omega_*$
such that it is expressed as
\begin{equation}
  \label{eq:momamg}
  \vec{m} = m \, [ \sin\chi \, ( \cos (\Omega_* \, t) \, \vec{e}_x + 
  \sin (\Omega_* \, t) \, \vec{e}_y ) + \cos\chi \, \vec{e}_z ]
\end{equation}
The inclination of the line of sight with respect to the rotational
axis, and defined by the unit vector $\vec{n}$, is denoted by~$\zeta$,
it lies in the $(Oyz)$~plane thus
\begin{equation}
 \vec{n} = \sin\zeta \, \vec{e}_y + \cos\zeta \, \vec{e}_z
\end{equation}
Moreover, the wind is expanding radially outwards at a velocity~$V$
close to the speed of light denoted by~$c$.

Our model involves some geometrical properties related to the magnetic
field structure and some dynamical properties related to the emitting
particles. Furthermore, in order to compute the light curves and the
corresponding spectra, we need to know the emissivity of the wind due
to inverse Compton scattering. This is explained in the next
paragraphs.

\subsection{Magnetic field structure}

\begin{figure}
 \includegraphics[scale=0.8]{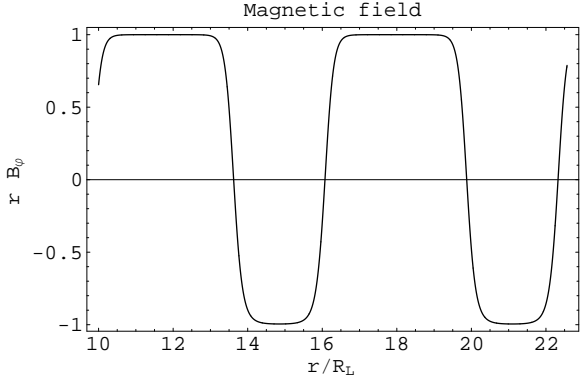}
 \caption{Structure of the azimuthal magnetic field in the equatorial
   plane at a given time. Note the smooth polarity reversal when
   crossing the current sheets. The scaling by a factor~$r$ removes
   the radial decrease of the amplitude and helps to better visualize
   the thickness of the current sheets. Distances are normalized to
   the radius of the light cylinder.}
 \label{fig:Bphi} 
\end{figure}

The striped wind has been introduced by \cite{1990ApJ...349..538C} and
revisited by \cite{1994ApJ...431..397M}. Here, we adopt a geometrical
structure of the wind based on the asymptotic magnetic field solution
given by \citet{1999A&A...349.1017B}. This magnetic field model,
although being only a crude approximation of the distant magnetic
field, is nevertheless very convenient because it furnishes a simple
analytical expression containing all the important features for an
oblique rotator (relativistically expanding current sheets).  The
observed geometry is a well-known consequence of retardation effects
winding up the stripes into a spiral. We emphasize that the striped
wind solution as described in \citet{1999A&A...349.1017B} is not a
necessary condition. A realistic model should lift some of the
assumptions made, like a constant expansion speed for instance or
infinitely thin current sheet. Nevertheless, the main characteristics
of the pulsed emission properties would be preserved.  However, in
this first attempt to model the high-energy emission, we work with
this simplified model. We leave improvements for future work. Thus,
outside the light cylinder, the magnetic structure is replaced by two
magnetic monopoles with equal and opposite intensity.  The current
sheet sustaining the magnetic polarity reversal arising in this
solution, expressed in spherical coordinates~$(r, \theta, \varphi)$ is
defined by
\begin{equation}
  \label{eq:Rs}
  r_s(\theta,\varphi,t) = \beta \, r_L \, \left[ 
  \pm \arccos ( - \cot\theta \, \cot\chi) + \frac{c\,t}{r_L} - 
  \varphi + 2\,l\,\pi \right]
\end{equation}
where $\beta=V/c$, $r_L=c/\Omega_*$ is the radius of the light
cylinder, $t$ is the time as measured by a distant observer at rest,
and $l$ an integer. Because of the ideal MHD assumption, this surface
is frozen into the plasma and therefore moves also radially outwards
at a speed~$V$.  Strictly speaking, the current sheets are infinitely
thin. However, as was already done for the study of the synchrotron
polarization of the pulsed emission, \cite{2005ApJ...627L..37P}, we
release this restrictive and unphysical prescription. Indeed, the
current sheet are assumed to have a given thickness, parameterized by
the quantity~$\Delta_\varphi$.  Moreover, inside the sheets, the
particle number density is very high while the magnetic field is weak.
In whole space, the magnetic field is purely toroidal and given by
\begin{equation}
  \label{eq:Bphi}
  B_\varphi = B_{\rm L} \, \frac{R_{\rm L}}{r} \, \eta_\varphi
\end{equation}
The strength of the magnetic field at the light-cylinder is denoted
by~$B_{\rm L}$. In the original work of \cite{1999A&A...349.1017B},
the function $\eta_\varphi$ is related to the Heaviside unit step
function and can only have two values $\pm 1$, leading to the
discontinuity in magnetic field when crossing the surface defined by
Eq.~(\ref{eq:Rs}). In order to make such transition more smooth, we
redefine this unit step function by something going smoothly from -1
to +1 (and conversely) prescribed by the following expression
\begin{eqnarray}
  \label{eq:etaphi}
  \eta_\varphi & = & \tanh( \Delta_\varphi \, \psi ) \\
  \psi & = & \cos \theta \, \cos \chi + \sin \theta \, \sin \chi \, 
  \cos \, [ \varphi - \Omega_* \, ( t - r / V ) ]
\end{eqnarray}
With this formula, the transition layer has a thickness of
approximately~$\Delta_\varphi$. The azimuthal dependence in $\varphi -
\Omega_* \, ( t - r / V ) $ represents a $m=1$ spiral wave propagating
radially outwards at a speed $V$ and rotating at the speed of the
neutron star $\Omega_*$. The structure of the azimuthal magnetic field
in the equatorial plane is shown in Fig.~\ref{fig:Bphi}

\subsection{Particle distribution function}

\begin{figure}
 \includegraphics[scale=0.8]{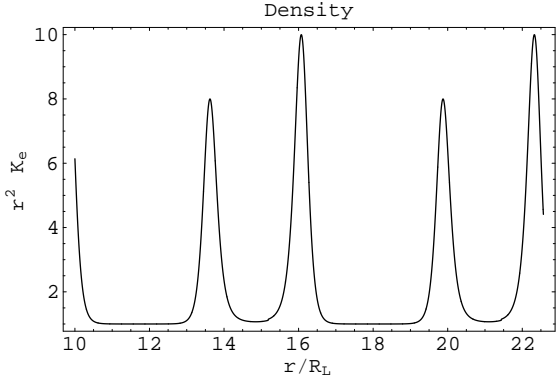}
 \caption{The radial variation of the particle density number in the
   cold magnetized part (low density) and in the current sheets (high
   density) when looking in the equatorial plane for a given time. The
   scaling with the factor~$r^2$ shows a constant density profile and
   a slightly different peak intensity in the current sheets.
   Distances are normalized to the radius of the light cylinder.}
\label{fig:Ke}
\end{figure}

The innermost regions of the pulsar magnetosphere is believed to be a
site of high-energy pair production feeding the wind with
ultrarelativistic electrons and positrons.  For these emitting
particles, which are very dense in the current sheets where the
magnetic pressure is weak, we adopt an isotropic distribution function
in momentum space in the comoving frame of the wind. It is given by a
power law in energy of index~$p$, with a sharp low and high-energy
cut-off, $\gamma_{min}$ and $\gamma_{max}$ respectively, such that the
particle number density at time~$t$ and position~$\vec{r}$ with energy
between~$\gamma$ and $\gamma+d\gamma$ is
\begin{equation}
  \label{eq:PartDistr}
  n_e(\gamma,\vec{r},t) \, d\gamma = K_e(\vec{r},t) \, \gamma^{-p} \, d\gamma \;\;\;
  \mathrm{with} \;\;\; \gamma_{min} \le \gamma \le \gamma_{max}
\end{equation}
$K_e(\vec{r},t)$ is related to the number density of emitting
particles in the current sheet. The particular form of the magnetic
field in the current sheets, decreasing like a tangent hyperbolic
function~$\tanh$ as given in Eq.~(\ref{eq:Bphi}), suggests to use a
plasma density profile dictated by the exact solution of the
relativistic Harris current sheet, namely a secant cosines hyperbolic
function~${\rm sech}$, see for instance \cite{2007PPCF...49.1885P}.
We thus adopt the following expression for the density
\begin{equation}
  \label{eq:Densite}
  K_e(\vec{r},t) = \frac{( N - N_0 ) \, 
    {\rm sech}^2 (\Delta_\varphi \, \psi) + N_0}{r^2}
\end{equation}
$N_0$ sets the minimum particle density in the magnetized and cold
plasma, i.e. between the current sheets, whereas $N$ defines the
highest density inside the sheets.  However, in order to allow
different peak intensity in the light curves, we choose different
maximum densities in two consecutive sheets, $N_1$ and $N_2$.  This
discrepancy could be explained by a different pair creation rate in
the two polar caps, feeding the stripes with different numbers of
particles. The radial motion of the wind at a fixed speed imposes an
overall $1/r^2$ dependence on this quantity, due to the conservation
of the number of particles. An example of particle density number,
multiplied by this $r^2$ factor for the Geminga pulsar is shown in
Fig.~\ref{fig:Ke}. Note, however, that adiabatic losses in the current
sheets due to pressure work will cool down this distribution function
in such a way that $K_e$ decreases like $1/r^{2(p+2)/3}$
\citep{1994plas.conf..225K}.  As already done in a previous work, we
assume the emission commences when the wind crosses the surface
defined by~$r = r_0 > r_{\rm L}$. Why should emission suddenly start
at this arbitrary radius $r_0$? If the stripes are responsible for the
observed pulsed emission, it corresponds to regions where the wind is
well form, i.e. outside the light-cylinder. Moreover, to explain the
highly temporal stability of these pulses, emission should occur
closest to the light cylinder because magnetic field is strongest
there. Indeed, we do not expect strong radiation from far away regions
because the particle density number in the current sheets decreases
like the square of the radius. Therefore, significant radiation is
emanating from the base of the wind close to the launching region,
probably near the light-cylinder.  Unfortunately, the transition
region between the closed magnetosphere and the wind regime is poorly
described, we only know that it should happen in the vicinity of the
light-cylinder. The transition region would smoothly switch from an
unpulsed component (in the closed magnetosphere) to a pulsed component
(in the wind).  In our model, we suppress this transition, making it
of size zero.  Thus, the relevant pulsed emission starts sharply when
crossing the shell $r=r_0$ . Adding a smooth transition will not
change the conclusions discussed in the next section. Note also that
the whole wind is radiating, any plasma element located at a distance
larger than $r_0$ will contribute to the total emissivity and
\textit{not only when particles cross} the shell $r_0$. However,
because of the fast decrease in intensity, only the first few stripes
provide almost all the pulsed component.

Let us now discuss the exact form of the emissivity functions in the
striped wind for the inverse Compton radiation.

\subsection{Inverse Compton emissivity}

Let us first recall the general expression for Inverse Compton
radiation of an electron embedded in an isotropic photon field. We
assume an isotropic distribution of mono-energetic target
photons~$\varepsilon$ with density $n_\gamma(\varepsilon)$ in the
observer frame.  The number of photons of energy~$\varepsilon_1$
scattered per unit time and per unit frequency by an
ultra-relativistic electron is
\begin{equation}
  \label{eq:Difcs}
  \frac{dN}{dt \, d\tilde{\varepsilon}_1} = \frac{3}{4} \, 
  \frac{\sigma_\mathrm{T} \, c}{\gamma^2 \, \tilde{\varepsilon}} \, f(q) \,
  n_\gamma(\tilde{\varepsilon}) \, d\tilde{\varepsilon}
\end{equation}
For convenience, energies of photons are scaled to the rest mass
energy of an electron~$m_e\,c^2$ such that $\varepsilon =
\tilde{\varepsilon} \, m_e \, c^2$ and $\varepsilon_{1} =
\tilde{\varepsilon}_{1} \, m_e \, c^2$. $\sigma_T$ is the Thomson
cross section and the function~$f$ is defined in
\cite{1970RvMP...42..237B} by
\begin{eqnarray}
  \label{eq:ICF}
  f(q) & = & 1 + q + 2 \, q \, \ln q - 2 \, q^2 + 
  \frac{(4 \, \gamma \, \tilde{\varepsilon} \, q)^2}
  {2 \, (1 + 4 \, \gamma \, \tilde{\varepsilon} \, q)} \, ( 1 - q ) \\
  q & = & \frac{\tilde{\varepsilon}_1}{4 \, \gamma \, \tilde{\varepsilon}
    \, ( \gamma - \tilde{\varepsilon}_1)}
\end{eqnarray}
The power spectrum density of inverse Compton emission is therefore
\begin{equation}
  \label{eq:Ie}
  P_\mathrm{ic}(\varepsilon_1) = \varepsilon_1 \, \frac{dN}{dt \, d\varepsilon_1}
  = \tilde{\varepsilon}_1 \, \frac{dN}{dt \, d\tilde{\varepsilon}_1}
\end{equation}
Summing the contribution from each lepton in the distribution function
Eq.~(\ref{eq:PartDistr}), the total emissivity in the observer frame
will be
\begin{equation}
 \label{eq:j_ic}
  j_\mathrm{ic}^\mathrm{obs}(\varepsilon_1) = \int_{\gamma_{min}}^{\gamma_{max}} \,
  \varepsilon_1 \, \frac{dN}{dt \, d\varepsilon_1} \, n_e^\mathrm{obs}(\gamma) \, d\gamma
\end{equation}
In the case of ultra-relativistic electrons, their rest mass energy is
negligible compared to their kinetic energy and thus they can be
treated as massless particles. Their energy (or equivalently their
Lorentz factor) then transforms according to the Doppler shift formula
for photons $\gamma = \mathcal{D} \, \gamma'$
\citep{2001ApJ...561..111G} where the Doppler factor associated with
the relativistic expansion of the wind $\vec{\beta}=\vec{V}/c$ with
Lorentz factor $\Gamma=1/\sqrt{1-V^2/c^2}$ is given by
\begin{equation}
 \mathcal{D} = \frac{1}{\Gamma \, ( 1-\vec{\beta} \cdot \vec{n})}
\end{equation}
$\vec{n}$ being an unit vector along the line of sight joining the
emitting element to the observer.  Moreover, the associated
distribution function expressed in the lab frame transforms according
to
\begin{equation}
  \label{eq:PartDistrObs}
  n_e^\mathrm{obs}(\gamma) = \mathcal{D}^2 \, n_e(\gamma/\mathcal{D})
  = K_e(\vec{r},t) \, \mathcal{D}^{p+2} \, \gamma^{-p} \;\;\; \mathrm{with} \;\;\;
  \mathcal{D} \, \gamma_{min} \le \gamma \le \mathcal{D} \, \gamma_{max}
\end{equation}
For the remaining of this paper, we will only consider the Thomson
regime for which $\gamma \, \tilde{\varepsilon} \ll 1$. The last term
on the RHS of Eq.(\ref{eq:ICF}) proportional to $(1-q)$ can thus be
dropped. In this limit, the integration of Eq.~(\ref{eq:j_ic}) can be
done analytically. To do this we need the following integral formula
\begin{eqnarray}
  \label{eq:IntThopIC}
  F_\mathrm{ic}(x) & = & \int x^{(p-1)/2} \, f(x) \, dx \nonumber \\
  & = & 2 \, x^{(p+1)/2} \,
  \left[ \frac{1}{p+1} - \frac{2\,x^2}{p+5} + \frac{2\,x\,\ln x}{p+3}
    + \frac{p-1}{(p+3)^2} \, x \right]
\end{eqnarray}
The emissivity can therefore be written as
\begin{equation}
  \label{eq:dfg}
  j_\mathrm{ic}^\mathrm{obs}(\varepsilon_1) = 2^{p-2} \, 3 \, \sigma_\mathrm{T} 
  \, c \, n_\gamma(\varepsilon) \, K_e \, \mathcal{D}^{ p+2} \, 
  \left( \frac{\varepsilon_1}{\varepsilon} \right)^{-(p-1)/2} \, 
  [ F_\mathrm{ic}(\mathrm{min}(x_1,1)) - F_\mathrm{ic}(\mathrm{min}(x_2,1)) ]
\end{equation}
where the upper and lower limits are respectively $x_1 = \varepsilon_1
/ 4 \, \gamma_1^2 \, \varepsilon$ and $x_2 = \varepsilon_1 / 4 \,
\gamma_2^2 \, \varepsilon$. the minimum and maximum Lorentz factor of
the leptons are $\gamma_1 = \mathcal{D} \, \gamma_{\rm min}$ and
$\gamma_2 = \mathcal{D} \, \gamma_{\rm max}$.  For frequencies well
above the lower cut-off and well below the upper cut-off, the total
intensity of inverse Compton scattering is then proportional to
\begin{equation}
  \label{eq:emisICObs}
  j_\mathrm{ic}^\mathrm{obs}(\varepsilon_1, \vec{r}, t) \propto
  K_e(\vec{r}, t) \, \varepsilon_1^{-(p-1)/2} \, 
  \mathcal{D}^{ p+2} \, n_\gamma(\varepsilon)
\end{equation}
Comparing the power index of the Doppler factor in Eq.~(3) of
\cite{2005ApJ...627L..37P} and Eq.~(\ref{eq:emisICObs}), the
dependence on~$p$ is different $s_{\rm sync} = (p+3)/2$ for
synchrotron and $s_{\rm ic} = (p+2)$ for inverse Compton. For $p>-1$,
beaming effect is weaker for synchrotron compared to inverse Compton
because $s_{\rm sync} < s_{\rm ic}$.  Therefore, the pulses in the
latter case will be more sharp than the pulses from the former.

\subsection{Light curves and spectra}

Knowing the inverse Compton emissivity, the light curves are obtained
by integration over the whole wind region. This wind is assumed to
extend from a radius~$r_0$ to an outer radius~$r_{\rm out}$ which can
be interpreted as the location of the termination shock.  However,
because of the fast decrease of particle density with radius, there is
no need to go such far in practice when performing the numerical
integration. Moreover, the radius $r_0$ is interpreted as the
transition region between the closed magnetosphere and the launch of
the wind. We adopt a sharp unrealistic transition by switching on the
emission when crossing this surface. However, in reality we would
expect emission also below this radius. But this region being poorly
understood, we start with this crude model. The shape of the pulses
will not be affected by this approximation, it is equivalent to
neglect some contribution from the DC component.  Therefore, the
inverse Compton radiation at a fixed observer time~$t$ is given by
\begin{eqnarray}
  \label{eq:CourbeLum}
  I_\mathrm{ic}^\mathrm{obs}(\omega, t) & = & 
  \int_{r_0}^{r_{\rm out}} \int_{0}^{\pi} \, \int_0^{2\pi}
  j_\mathrm{ic}^\mathrm{obs}(\omega, \vec{r}, t_{\rm ret}) \, r^2 \, 
  \sin\theta \, dr \, d\theta \, d\varphi
\end{eqnarray}
The retarded time is expressed as $t_{\rm ret} = t - ||\vec{R}_0 -
\vec{r}||/c \approx t - R_0/c + \vec{n} \cdot \vec{r} / c$. The
approximation is valid if the observer, located at $\vec{R}_0$, is
very far away from the radiating system, $R_0\gg r_{\rm out}$.
Eq.(\ref{eq:CourbeLum}) is integrated numerically. We compute the
inverse Compton intensity for several frequencies from far below the
low cut-off frequency to far over the high frequency cut-off. We are
therefore able to predict the phase resolved spectral variability and
the pulse shape simultaneously.  The results and applications to the
Geminga pulsar which is known to be a $\gamma$-ray pulsars are
discussed in the next section.

\section{APPLICATION TO $\gamma$-RAY PULSARS}
\label{sec:Application}

We apply the aforementioned model to inverse Compton scattering of
low-energy photons from the cosmic microwave background with typical
energy of $\varepsilon_{CMB} = k_B\,T_\mathrm{CMB} =
2.36\times10^{-4}$~eV and energy density of $2.65 \times
10^5~\mathrm{eV/m}^3$ by the ultrarelativistic leptons flowing in the
current sheets of the striped wind. The CMB radiation field is assumed
to be isotropic. We focus on the Geminga pulsar for which the useful
parameters are listed in Table~\ref{tab:Parametres}.

An important parameter of our model is the maximal Lorentz factor
reached by pairs. Indeed, it directly determines the cut-off energy in
the pulsed spectra. Note also that the minimal Lorentz factor does not
play a significant role to fit the observations. We can find an order
of magnitude of $\gamma_{\rm max}$ in the following manner.  Assuming
that the highest energy emanates from the inverse Compton scattering
with the CMB, we can deduce the maximal Lorentz factor of the
electrons in the observer frame.  Indeed, no pulsed emission is seen
above roughly~$30$~GeV \citep{2004IAUS..218..399T}. The cut-off energy
$\varepsilon_{cut}$ can be fixed at about 3~GeV. The maximum electron
Lorentz factor is therefore obtained by
\begin{equation}
  \label{eq:MaxLorFact}
  \gamma_{\rm max}^{\rm obs} = 
  \sqrt{\frac{\varepsilon_{cut}}{4\,\varepsilon_{CMB}}}
\end{equation}
Thus a typical value is $\gamma_{\rm max}^{\rm obs} \approx 10^{6.5}$.
The Lorentz factor of the wind, assumed to be constant, according to
previous results \cite{2005ApJ...627L..37P}, can be set to a few tens.
Adopting this value, the maximum Lorentz factor in the particle
distribution function Eq.~(\ref{eq:PartDistr}) in the comoving frame
is typically~$\gamma_{\rm max} \approx 10^5$.

\begin{table*}[htbp]
  \centering
  \begin{tabular}{cccccccc}
    \hline
    Pulsar  & $P$ (ms) & d (kpc) & $\log(B)$ (T) & 
    $B_L$ (T) & $R_L$ (km) & $\zeta$ (degree) & $\chi$ (degree) \\
    \hline
    Geminga & 237.09  &  0.15  &  8.21  &  0.112  & 11312.4 & 90 & 60 \\
    \hline
  \end{tabular}
  \caption{Properties of Geminga pulsar 
    assuming $R_*=10$~km. Period, distance and magnetic field strength are
    taken from \cite{1993ApJS...88..529T}.}
  \label{tab:Parametres}
\end{table*}


In our best fit, we choose an inclination of the magnetic moment with
respect to the rotation axis of $\chi=60^o$. In order to obtain a
phase separation of 0.5 between the two pulses, we have to adopt an
inclination of the line of sight $\zeta=90^o$. The Lorentz factor of
the wind is $\Gamma=10$.

\begin{figure}[htbp]
  \centering \includegraphics[scale=0.5]{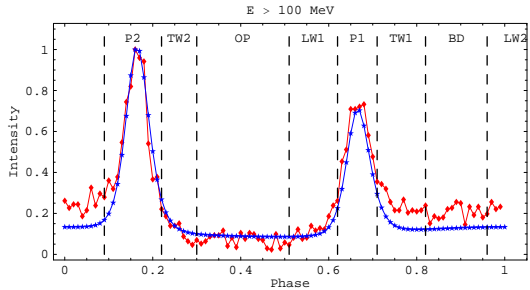}
  \caption{Gamma-ray light curve above 100~MeV of Geminga fitted with the
    inverse Compton emission from the striped wind.}
  \label{fig:Geminga_cl}
\end{figure}

\begin{figure}[htbp]
  \centering \includegraphics[scale=0.8]{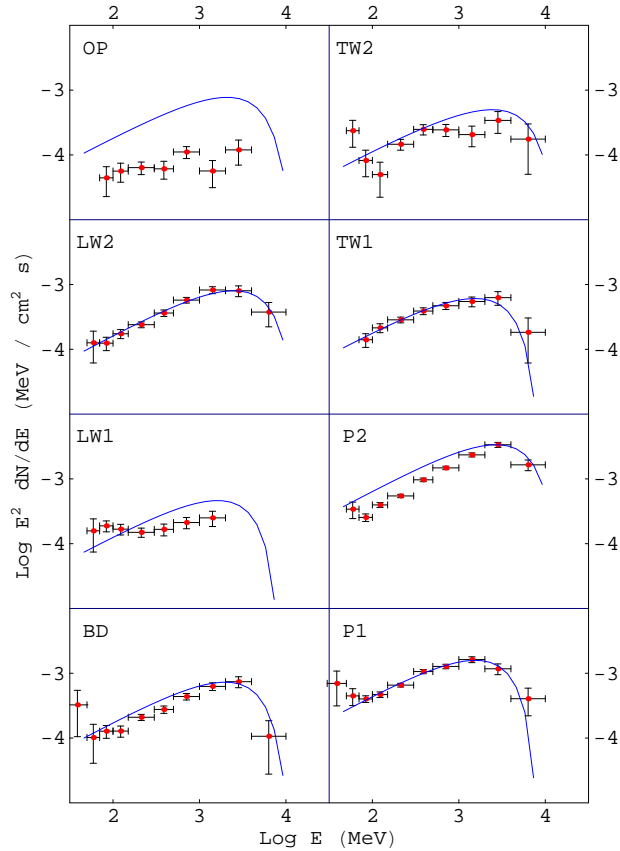}
  \caption{Phase-resolved inverse Compton emission from 
    the Geminga pulsar for different phase intervals: bridge (BD),
    leading wing~1/2 (LW1/LW2), off-pulse (OP), peak~1/2 (P1/P2),
    trailing wing~1/2 (TW1/TW2). Exact definition of these intervals
    and data are taken from \cite{1998ApJ...494..734F}.}
  \label{fig:Geminga_ph}
\end{figure}

Results for the light-curve above 100~MeV and the definition of the
different phase intervals are shown in Fig.~\ref{fig:Geminga_cl}.  The
rising and falling shape of both pulses are well fitted by our model.
In one wavelength of the striped wind, or equivalently in one period
of the pulsar, there are two current sheets which give rise to strong
emission because of the high-density in it. This explains the double
pulse structure which is observed at any frequency. Moreover, the two
pulses are very similar in shape, with only a difference in maximum
intensity. We explain this discrepancy by having slightly different
particle density numbers within the current sheets. The corresponding
phase-resolved spectra are shown in Fig.~\ref{fig:Geminga_ph}. The
spectral variability is reproduced with satisfactory accuracy except
for the OP phase for which the intensity is overestimated.

The intensity and the width of both peaks is matched with good
accuracy. Recall that the width is strongly influenced by the size of
the current sheets and the Lorentz factor of the wind, due to
relativistic beaming effects. For P1, the fitted spectrum lies within
the error bars of the EGRET data whereas for P2, the slope is slightly
overestimated.  Nevertheless, the shape of the cut-off in both cases
is retrieved.  An excellent fit is also found for the trailing wing~1
(TW1) and the leading wing~2 (LW2). The bridge (BD) and the trailing
wing~2 (TW2) follow the same trend. The cut-off although apparent in
some phase intervals does not clearly show up in the off-pulse (OP) or
in the leading wing~1 (LW1) interval.  Observations form the future
GLAST satellite will give useful information in this energy band and
help to discriminate between different pulsar emission models.

\section{CONCLUSION}
\label{sec:Conclusion}

In the striped wind model, the pulsed high-energy emission from
pulsars arises from regions well outside the light-cylinder. The
geometry of the wind does not depend on the precise magnetospheric
structure close to the neutron star surface. Therefore, a detailed
model of the physics in the inner pulsar magnetosphere is not required
to make robust predictions concerning radiation emanating from the
striped wind.  By computing the inverse Compton emission on the CMB
photons in the Thomson regime, we were able to fit the EGRET data of
the light-curves as well as the phase-resolved spectra for at least
one gamma-ray pulsar, like Geminga.

Measuring the polarization properties of this high-energy emission
would provide a valuable diagnostic of the physical process
responsible for these MeV-GeV photons. In the future, experiments are
planned to measure the polarization in X-rays or gamma-rays, like the
PoGOLite satellite \citep{2007arXiv0709.1278K}.

Because the radiation is mostly emanating from the basis of the
striped wind where the density of particle is highest, a better
description of the transition between the closed magnetosphere and the
wind is required to obtain a detailed view of the magnetic field
configuration.

Note also that the same kind of computation can be performed for the
Crab pulsar for which we know the polarization properties in optical
bands and can therefore predict the polarization at higher energies by
extrapolation.  

\begin{acknowledgements}
  I am grateful to John Kirk for helpful discussions and suggestions
  and to Dave Thompson for sending me the EGRET data.  This work was
  partly supported by a grant from the G.I.F., the German-Israeli
  Foundation for Scientific Research and Development.
\end{acknowledgements}

\bibliographystyle{/home/petri/texmf/tex/latex/aa-package/bibtex/aa}

\end{document}